\documentclass[conference]{IEEEtran}
\IEEEoverridecommandlockouts
\usepackage{cite}
\usepackage{amsmath,amssymb,amsfonts}
\usepackage{algorithmic}
\usepackage{graphicx}
\usepackage{textcomp}
\usepackage{import}
\usepackage{placeins}
\usepackage{textcomp}
\usepackage{url}
\usepackage[hypertexnames=false,bookmarks=false]{hyperref} 
\hypersetup{colorlinks,linkcolor={red},citecolor={blue},urlcolor={black}} 
\usepackage[euler]{textgreek} 
\usepackage{blindtext}
\usepackage{ragged2e}
\usepackage[table,xcdraw]{xcolor}
\usepackage{tabularx}
\usepackage{multirow} 
\usepackage{comment}
\usepackage{stfloats}

\newcolumntype{s}{>{\hsize=.5\hsize}X}
\newcolumntype{q}{>{\hsize=.25\hsize}X}

\ifCLASSOPTIONcompsoc
    \usepackage[caption=false, font=normalsize, labelfont=sf, textfont=sf]{subfig}
\else
\usepackage[caption=false, font=footnotesize]{subfig}
\fi

\def\BibTeX{{\rm B\kern-.05em{\sc i\kern-.025em b}\kern-.08em
    T\kern-.1667em\lower.7ex\hbox{E}\kern-.125emX}}

\IEEEaftertitletext{\vspace{-0.7\baselineskip}}
\begin{document}
\bstctlcite{IEEEexample:BSTcontrol}

\title{Neural implants and human safety: single-fault detection for DC-coupled recording front ends
\vspace{-6mm}

}

\author{%
\IEEEauthorblockN{Dimitris Antoniadis$^{1,2}$, Timothy G. Constandinou$^{1,2,3}$}
\IEEEauthorblockA{$^{1}$Dept. of Electrical \& Electronic Engineering, Imperial College London, South Kensington Campus, London, SW7 2AZ, UK}
\IEEEauthorblockA{$^{2}$UK Dementia Research Institute Centre for Care Research \& Technology at Imperial College London \& University of Surrey}
\IEEEauthorblockA{$^{3}$Mint Neurotechnologies Ltd, 125 Wood Street, London, EC2V 7AW, UK}
\IEEEauthorblockA{Email: \{dimitris.antoniadis20, t.constandinou\}@imperial.ac.uk}
}

\IEEEaftertitletext{\vspace{-3mm}}
\maketitle

\begin{abstract}
DC-coupled analogue front ends  (AFEs) for neural implants provide a low-area solution. However, removing the coupling capacitor eliminates the intrinsic barrier that protects cortical tissue: a single-fault event, such as gate-oxide breakdown of a low-noise amplifier (LNA) input transistor, can open a direct DC path from the supply rail into the brain. On the stimulation side this hazard is well understood, and single-fault tolerance is enforced by a series DC-blocking capacitor; on the recording side, DC-coupled front ends discard the equivalent safeguard, yet their protection has gone almost unexamined. This paper presents a single-fault detection mechanism that monitors the LNA for the DC imbalance produced by such a failure and disables the amplifier before the resulting fault current can irreversibly damage tissue. The imbalance is encoded in the duty cycle of a current-starved relaxation oscillator and read out as a time-to-digital measurement. Designed in 65\,nm, the mechanism 
resolves a worst-case fault of 6.4~nA across all corners within 0.81~ms\,---\,compliant with the ISO~14708-3 limit for an 8533\,\textmu m\textsuperscript{2} electrode\,---\,opening a broader discussion of safety in DC-coupled recording.
\end{abstract}

\section{Introduction} \label{sec:Introduction}

Neural implants have emerged as a transformative class of devices, forming a link between the nervous system and external digital systems~\cite{mcfarland2008emulation, wolpaw2020brain, gonzalez2024bioelectronic}. They span several roles. Recording interfaces decode neural activity for communication and control, neuromodulation devices deliver therapeutic stimulation such as deep-brain stimulation for Parkinson's disease, and closed-loop devices both sense and stimulate, as in responsive neurostimulation for epilepsy~\cite{brunner2015bnci, robinson2021emerging, lerman2025next}. This progress is increasingly driven by distributed, minimally invasive mm-scale implants, which push channel counts higher and enable new classes of application~\cite{ozbek2026, musk2019integrated, ahmadi2019towards, rapeaux2021implantable, Szostak2020, muller2021miniaturized}.

These implants acquire neural signals whose class is set by electrode scale. Small penetrating microelectrodes resolve both extracellular action potentials (EAPs), spanning 50\,--\,500\,\textmu V\textsubscript{pp} across 100\,Hz\,--\,10\,kHz, and local field potentials (LFPs), spanning 0.5\,--\,5\,mV\textsubscript{pp} across 100\,mHz\,--\,200\,Hz, whereas larger macroelectrodes are limited to LFPs and do not reliably resolve single units~\cite{kandel2000principles,mollazadeh2009wireless,gosselin2011recent,white2010real,patil2008development}. Acquisition is handled by an analogue front end (AFE) in which an array of low-noise amplifiers (LNAs) feeds either a shared multiplexed analogue-to-digital converter (ADC) or, in direct-digitisation designs, a per-channel ADC~\cite{antoniadis2026, villa2024enhancing, woeppel2017recent, campbell2018chronically}.

\begin{figure}[!b]
    \centering
    \centerline{\includegraphics[width=1.0\columnwidth]{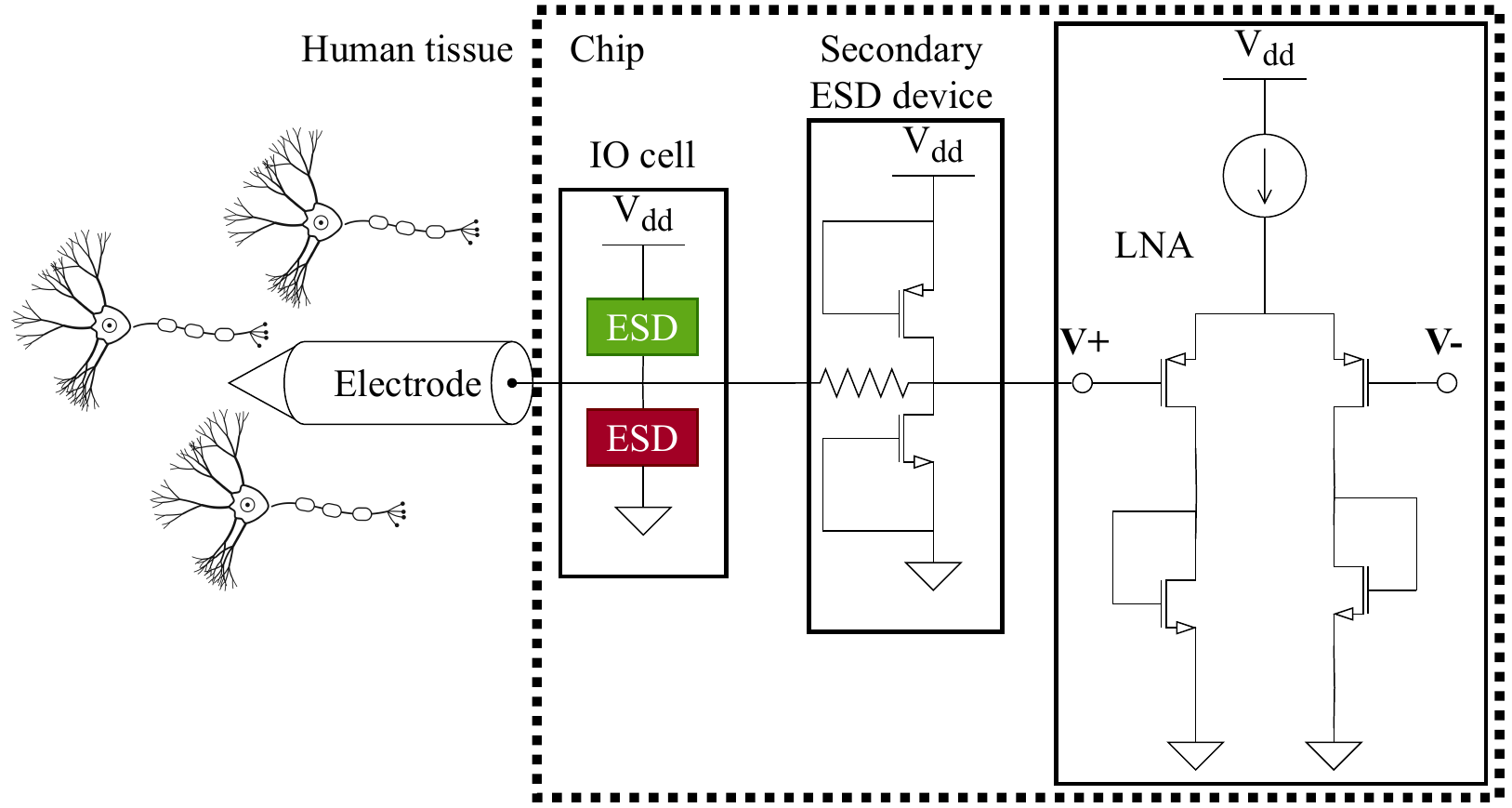}}
    \caption{Simplified schematic of the recording path, from electrode through the ESD devices to the LNA. Three failure points are identified: the IO-cell ESD, the secondary ESD device and the LNA input transistor. On failure, any of the three opens a direct DC path from the supply rail into brain tissue.}
    \label{fig:simplified_signal_path}
\end{figure}

A defining constraint of the AFE is the large, time-varying DC offset that arises at the electrode–tissue interface of miniaturised electrodes~\cite{harrison2003low}. The dominant historical solution has been the AC-coupled AFE, in which a series coupling capacitor rejects the electrode DC offset and blocks any DC path from the front-end electronics into cortical tissue~\cite{bagheri2016low,muller2021miniaturized}. ISO 14708-1:2014 (Clause 19.3) requires that the failure of any single component, part or software program shall not cause an unacceptable hazard, and on the stimulation side the series capacitor satisfies this single-fault requirement~\cite{iso14708-1}. The blocking capacitor is therefore both a signal-conditioning and a safety element. The cost of this safety is area. Realising the sub-hertz high-pass corner required to capture LFPs demands coupling capacitors that are prohibitively large, and this footprint scales with channel count, directly opposing the drive toward high-density. For this reason DC-coupled AFEs are increasingly preferred because, by removing the coupling capacitor they achieve a far smaller per-channel area~\cite{bagheri2016low}. However, the DC-coupled AFEs inherit a liability as there is no longer an intrinsic barrier to DC current flow between the front-end supply and the electrode. A single-fault event such as a gate-oxide breakdown of an LNA input transistor can open a direct DC path from the supply rail through the electrode and into the cortical tissue. A simplified schematic of a DC-coupled front end is shown in Fig.~\ref{fig:simplified_signal_path}. The bio-signal reaches the gates of the LNA differential input pair through the pad and the ESD devices.

Despite progress in stimulation safety and electrode characterisation, DC-coupled AFEs cannot yet guarantee patient safety during recording, which has limited the clinical and commercial uptake of neural implants~\cite{merrill2005electrical,cogan2008neural,cogan2016tissue,antoniadis2026}. Limited work addresses recording-side protection for these architectures, available only as a handful of patents using current-limiting techniques such as RC circuits~\cite{US6996435,goyal2021development}.

A further hazard lies in the electrostatic discharge (ESD) protection itself: the device at the pad and the secondary device between the pad and the input-transistor gate. If either fails, a direct DC path can form from the supply rail into brain tissue. The pad ESD sits between the electrode and the coupling capacitor, so this route bypasses the DC-blocking capacitor and remains a hazard even in AC-coupled front ends. Existing ESD designs for biosignal amplifiers~\cite{US8560041B2, US9214799B2} target circuit protection, not human safety.



In parallel, regulation of neural recording and stimulation devices is evolving, setting strict limits to ensure human safety. ISO 14708-3:2017, although focused on stimulation, limits the DC current density at the electrode to 0.75\,\textmu A/mm\textsuperscript{2} (Clause 16.2)~\cite{iso14708}. No equivalent limit governs the recording side. 


This paper presents a single-fault detection mechanism that monitors the LNA for the signature of an internal device failure and disables it before the resulting fault current can damage tissue. The signature is encoded ratiometrically in the duty cycle of a current-starved relaxation oscillator, making detection robust to process variations. The minimum electrode tip area satisfying the ISO~14708-3 limit is derived as 8533\,\textmu m\textsuperscript{2}, and the grace period before the water window is reached is quantified for a SIROF electrode characterised in the literature~\cite{lutz2025analysis}.

Section~\ref{sec:AFE_architecture} details the proposed LNA architecture and its single-fault detection, Section~\ref{sec:results} reports the simulation results, and Section~\ref{sec:conclusion} concludes and outlines future work.
 \begin{figure}[!t]
    \centering
    \centerline{\includegraphics[width=1.0\columnwidth]{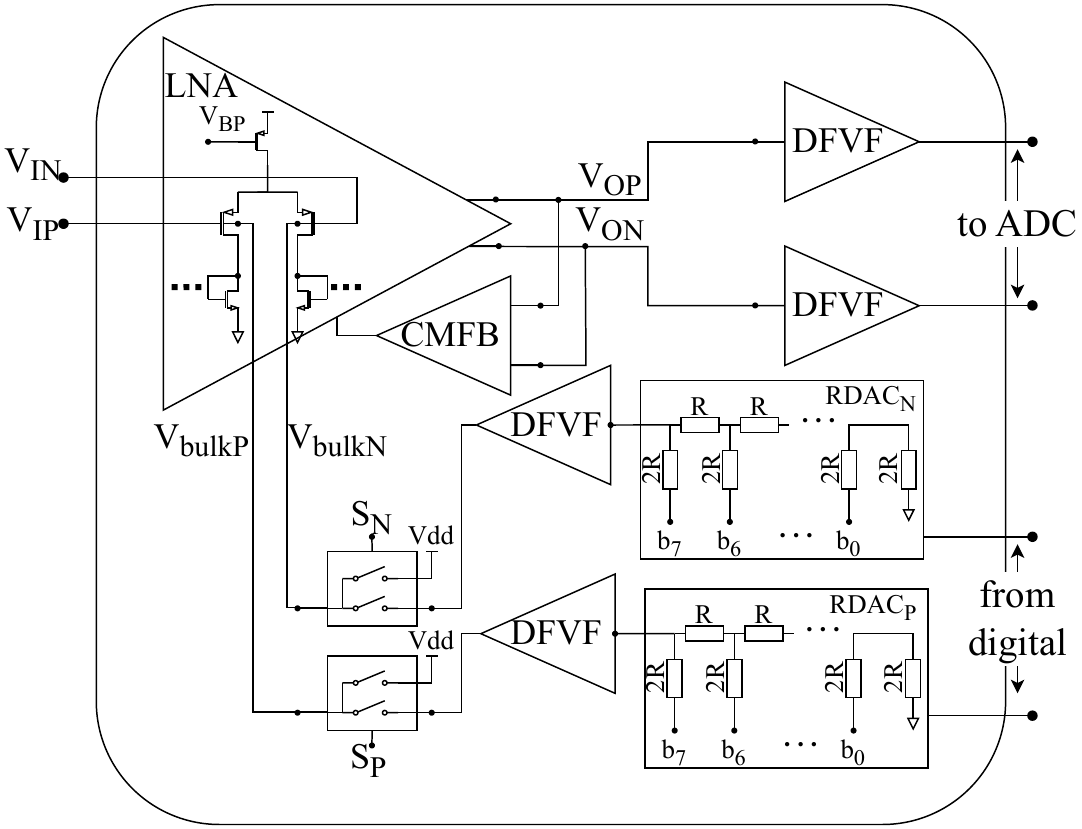}}
    \caption{Low noise amplifier and complementary circuits of the AFE architecture per channel~\cite{antoniadis2026}.}
    \label{fig:LNA}
\end{figure}  
\section{Single-Fault Tolerance Architecture} \label{sec:AFE_architecture}
\subsection{AFE Architecture Overview}
\begin{figure*}[!h]
    \centering
    \includegraphics[width=\textwidth]{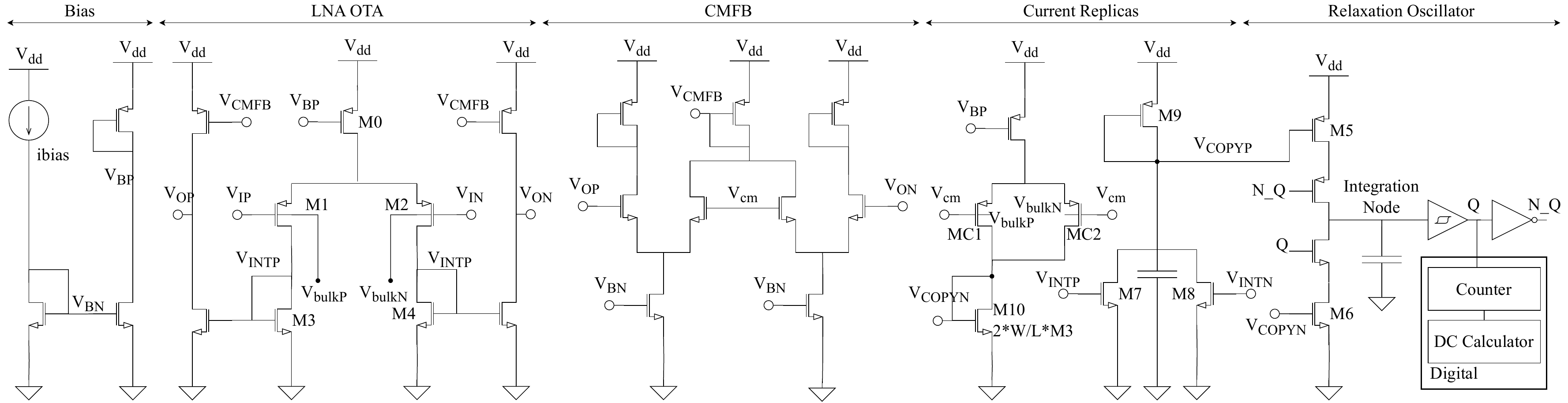}
    \caption{Simplified LNA schematic and fault detection mechanism using a relaxation oscillator.}
    \label{fig:simplified_sch_ota}
\end{figure*}

The single-fault tolerance detection mechanism presented in this paper enhances the AFE presented in \cite{antoniadis2026}. The AFE includes a fully differential, two stage, current mirror amplifier. It also includes a common feedback amplifier to set the differential output of the amplifier around a common mode voltage. The DC electrode offset and the low frequency components are rejected through a DC servo loop employing a digital low pass filter which controls the bulk of the differential pair transistors through two pseudoresistor digital-to-analogue converters (RDACs). A number of differential flipped voltage followers (DFVF) are used for voltage buffering. The LNA architecture is shown in Fig.~\ref{fig:LNA}. 
\begin{figure}[!t]
    \centering\centerline{\includegraphics[width=0.99\columnwidth]{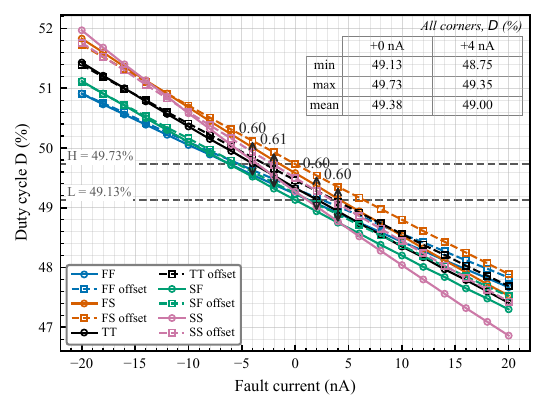}}
    \caption{Duty cycle and current fault detection across TT, FF, SS, SF, FS at 36\,$^\circ$C. Horizontal limits H and L account for worst case detection current of 6.4\,nA.}
    \label{fig:fault_vs_dc}
\end{figure}
\begin{figure}[!t]
    \centering\centerline{\includegraphics[width=0.99\columnwidth]{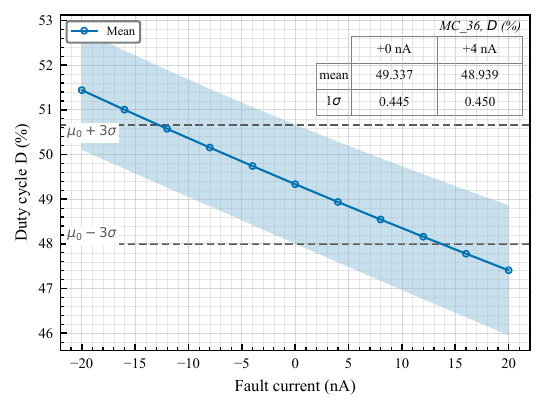}}
    \caption{Duty cycle and current fault detection across Monte Carlo simulation with mismatch and process variation at 36\,$^\circ$C. The blue line depicts the mean value and the light blue shape represents the  3\,$\sigma$ spread.}
    \label{fig:fault_vs_dc_MC}
\end{figure}

\subsection{LNA Architecture and Single-Fault Tolerance Circuit.}

The proposed LNA and its fault detection mechanism is shown in Fig.~\ref{fig:simplified_sch_ota}. The main sensing mechanism of the LNA is the input differential pair. In case of a device malfunction (M1 or M2) such as gate-oxide breakdown, a direct path is formed from the source or the bulk or the drain to the gate and then to the electrode and the human tissue. As long as the LNA bias current of M0 is equal to the sum of the current mirror branches (I\textsubscript{M3}\,+\,I\textsubscript{M4}), there is no current leaking to the tissue or coming from the tissue. Therefore, an accurate comparison of these currents could indicate whether a malfunction exists.

The fully differential current mirror amplifier of the LNA is biased by a resistorless beta multiplier. During monte carlo simulations with process and mismatch variation at 25\,$^\circ$C, 36\,$^\circ$C, 47\,$^\circ$C, the average current value is 123\,nA and the standard deviation is 7.42\,nA~\cite{oguey1997cmos}. Therefore, accurate operations of absolute current addition or subtraction for comparison in the order of a few nano amperes can be difficult. Further to this the current at the current mirror branches M3 and M4 is governed by frequency of the sensed bio-signals which ranges from sub-Hz to around 10\,kHz. 

In order to accurately compare the bias current against the sum of the current mirror branches, a measurement is needed that is independent of process and rejects the high frequencies due to input bio-signal. 

The rejection of high-frequency spikes can be achieved by copying the current from M3 and M4 to M7 and M8 and integrating their addition over a capacitor. This integration extracts the DC component of the current (I\textsubscript{M3}\,+\,I\textsubscript{M4}) through a diode connected transistor M9. The bias current of M0 can be generated by a replica circuit of the first stage of the amplifier with differential input pair biased at common mode voltage and its current added over a diode connected transistor M10. A plain bias current mirror could replace the first-stage replica, but would not track the LNA first-stage current as the bulk voltage changes due to DC offset cancellation through bulk. The two replica current mirror circuits are shown in Fig.~\ref{fig:simplified_sch_ota}.

Having copied the (I\textsubscript{M3}\,+\,I\textsubscript{M4}) dc current and generated a replica of M0 bias current, these two currents should have the same integration time over a capacitor regardless of the capacitance value. A relaxation oscillator suits this operation. Current I\textsubscript{M9} and I\textsubscript{M10} are copied over to M5 and M6 respectively. The proposed relaxation oscillator uses a current starved schmitt trigger comparator and as a result the integration range is the same for both the charging phase and the discharging phase. Since the currents copied at M5 and M6 are equal, the oscillator should have approximately 50\,\% duty cycle. If the capacitor is large enough and the current is small enough, a counter in the digital can define the duty cycle of the oscillator with great accuracy forming a time-to-digital converter. The two capacitors used in Fig.~\ref{fig:simplified_sch_ota} are of total 20.86\,pF each and their size is a 10\,$\times$\,10 array of 10\,\textmu m\,*\,10\,\textmu m mimcap unit. Upon detection of a duty cycle that is out of limit the LNA can be disabled and a power switch should disable its supply.

\section{Results} \label{sec:results}

The proposed architecture was designed in 65\,nm technology. The design is estimated to maintain  similar footprint by having improved area utilisation compared to \cite{antoniadis2026}, covering area approximately equal to 0.174\,mm\textsuperscript{2}. This section presents the simulated results across all corners (TT, FF, SS, SF, FS) at 36\,$^\circ$C and Monte Carlo (MC) with mismatch and process variation. 

Fig.~\ref{fig:fault_vs_dc} shows the corner results of the proposed architecture. The presented results include simulations of a 1\,mV sinusoidal input signal of 3\,kHz. Simulation has been performed twice, once with 0\,V offset and once with 10\,mV offset where the high-performance (HP) mode for offset cancellation is enabled. For 0\,A fault current, the duty cycle ranges from 49.13\,\% to 49.73\,\%. It is shown that for approximately $\pm$10\,nA the difference between min and max at a specific fault current value is maintained stable around 0.6\,\%. 
If the lower threshold of detection is set to 49.13\,\% (L) and the
upper to 49.73\,\% (H)\,---\,a separation window of 0.6\,\%\,---\,the protection
mechanism could prevent worst-case faults from $-6.2$\,nA to 6.4\,nA. This means that ISO 14708-3:2017 allows the use of this architecture with an electrode that has geometric surface area (GSA) equal to 8533\,\textmu m\textsuperscript{2}.
\begin{figure}[!h]
    \centering\centerline{\includegraphics[width=0.99\columnwidth]{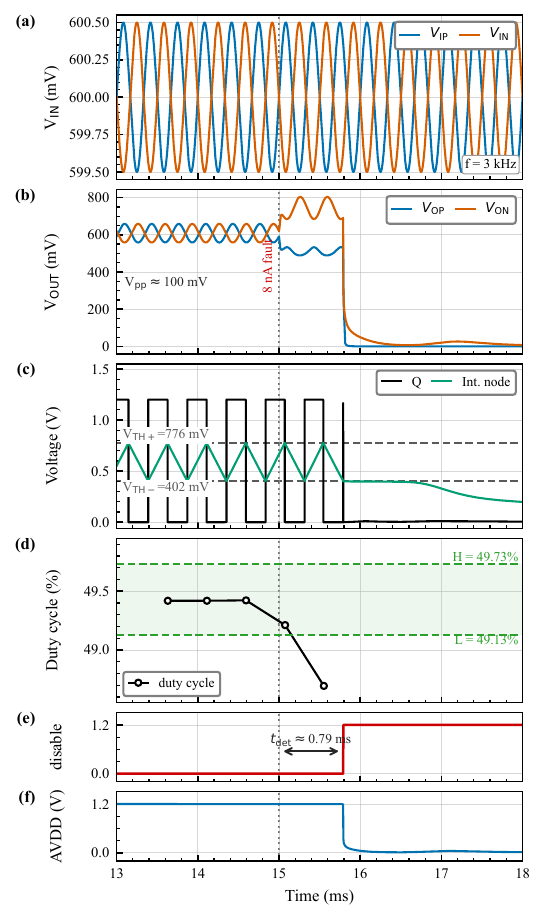}}
    \caption{Single-fault tolerance mechanism functionality for 8\,nA DC current: (a) sinusoidal input of 1\,mV and 3\,kHz; (b) output of the LNA; (c) voltage at the integration node of the relaxation oscillator generated by the detection mechanism current mirrors and the output signal Q of the oscillator (d) duty cycle calculation (e) disable LNA signal (f) LNA supply for TT at 36\,$^\circ$C. }
    \label{fig:osc}
\end{figure}

The oscillator frequency ranges from 1.23\,kHz to 3.06\,kHz across corners. With a trivial 100\,ns clock, the worst-case counter accuracy at 1.23\,kHz is 0.0123\,\%, introducing a negligible $\pm$0.02504\,\% relative error at the ideal detection thresholds, resolving the lower limit to 49.118 – 49.142\,\% and the upper limit to  49.718 – 49.742\,\%\,---\,far finer than required. At the fastest corner (3.06\,kHz) the same clock gives a coarser 0.0306\,\% resolution, still negligible against the 0.6\,\% separation window, while a single cycle completes in 0.33\,ms rather than the 0.81\,ms slow-corner period.

Fig.~\ref{fig:fault_vs_dc_MC} shows the Monte Carlo results of the proposed architecture. Results are slightly worse compared to the corner results which is expected due to mismatch. The average duty cycle is 49.337\,\%. The duty cycle for 0\,nA and 3$\sigma$ spread ranges from 48.003\,\% to 50.672\,\%. While 3$\sigma$ covers 99\,\% of the cases, in extreme cases it could account for fault currents of up to around 25\,nA, indicating that binning process can be applied during testing to select better quality chips or a calibration method applied so that, after binning or trimming, the architecture meets the ISO 14708-3 limit for the 8533 µm² electrode derived from the corner results.

Fig.~\ref{fig:osc} presents the functionality of the proposed circuit alongside the oscillator for 1\,mV peak to peak and 3\,kHz sinusoidal input. The integration node oscillates around the schmitt trigger limits. At 15\,ms of the transient simulation 8\,nA fault error is inserted. The duty cycle drops and triggers the kill switch of the LNA supply. Since the fault can be injected at an arbitrary point in the oscillator cycle, in the worst case two cycles are required to capture it. Actively pulling the nodes down, rather than letting them discharge passively, could be a future improvement.

A typical sputtered iridium oxide film (SIROF) electrode characterised in literature has GSA of the electrode tip equal to 2000\,\textmu m\textsuperscript{2}~\cite{lutz2025analysis}. 
The electrode is assumed to follow a simplified Randles model: a
double-layer capacitance (C\textsubscript{dl}) in parallel with a
resistor representing faradaic current. Should the LNA suffer a
device malfunction at the differential-pair input, the resulting
DC current follows the least-resistance path and charges
C\textsubscript{dl}. The SIROF electrode sits at a rest potential of
0.1\,--\,0.3\,V Ag\textbar AgCl~\cite{cogan2008neural,cogan2009sputtered}.
The water window before electrolysis happens is from -0.6\,V to 0.8\,V. Therefore, the worst case scenario would be charging the C\textsubscript{dl} from 0.3\,V to 0.8\,V Ag\textbar AgCl~\cite{cogan2008neural,cogan2009sputtered}. The capacitance of the electrode changes depending on the voltage of the electrode and for a mid range value, the C\textsubscript{dl} capacitor is equal to 1.14\,\textmu F\cite{lutz2025analysis}. The worst case detection current in Fig.~\ref{fig:fault_vs_dc} is equal to 6.4\,nA. Therefore grace time period available before fatal tissue damage for this SIROF electrode  would be in the order of \texttau\, = \textDelta V*C/I = 89.06\,s. As this electrode is smaller than the one derived from the corner results, its capacitance — and hence this grace time — is conservative. Given the worst case oscillator frequency being around 1.23\,kHz and irrespective of the grace time period, the detection circuit could detect the malfunction within worst case time of 0.81\,ms.

\begin{table}[!t]
\centering
\caption{\textsuperscript{*}LNA Performance in Low-Power Mode}
\label{tab:tabla_lp_mode}
\begin{tabular}{lccccc}
\hline
\textbf{Metric} & \textbf{Min} & \textbf{Max} & \textbf{Mean} & \textbf{MC Mean} & \textbf{MC $\sigma$} \\
\hline
$I$ (\textmu A) & 2.79 & 5.23 & 3.77 & 3.64 & 0.275\\
$f_c$ (kHz) & 7.72 & 12.7 & 10.03 & 10.11 & 0.765 \\
Gain (dB) & 39.74 & 40.6 & 40.22 & 40.11 & 0.444 \\
Noise$_\text{LF}$ (\textmu V\textsubscript{rms}) & 4.094 & 4.84 & 4.46 & 4.455 & 0.094 \\
Noise$_\text{HF}$ (\textmu V\textsubscript{rms}) & 9.28 & 12.2 & 10.77 & 10.73 & 0.449 \\
\hline
\end{tabular}
{\footnotesize\raggedright\textsuperscript{*}LNA with all circuits shown in Fig.~\ref{fig:LNA} except the 2 DFVF to ADC. Noise$_\text{LF}$: 0.1--200\,Hz. Noise$_\text{HF}$: 200\,Hz--10\,kHz.\par}
\end{table}

\begin{table}[t]
\centering
\caption{\textsuperscript{*}LNA Performance in Offset-Cancellation Mode}
\label{tab:tabla_hp_mode}
\begin{tabular}{lccccc}
\hline
\textbf{Metric} & \textbf{Min} & \textbf{Max} & \textbf{Mean} & \textbf{MC Mean} & \textbf{MC $\sigma$} \\
\hline
$I$ (\textmu A) & 4.23 & 7.733 & 5.70 & 5.6 & 0.427 \\
$f_c$ (kHz) & 13.91 & 19.1 & 16.16 & 17.08 & 0.997 \\
Gain (dB) & 41.52 & 42.06 & 41.84 & 41.78 & 0.268 \\
Noise$_\text{LF}$ (\textmu V\textsubscript{rms}) & 6.501 & 11.31 & 7.04 & 7.09 & 1.08 \\
Noise$_\text{HF}$ (\textmu V\textsubscript{rms}) & 9.33 & 12.52 & 11.06 & 14.95\textsuperscript{$\dagger$} & 12.4\textsuperscript{$\dagger$} \\
\hline
\end{tabular}
{\footnotesize\raggedright\textsuperscript{*}LNA with all circuits shown in Fig.~\ref{fig:LNA} except the 2 DFVF to ADC. Noise$_\text{LF}$: 0.1--200\,Hz. Noise$_\text{HF}$: 200\,Hz--10\,kHz.  \textsuperscript{$\dagger$}Distribution heavy-tailed due to extreme outliers; median $\approx 11$\,\textmu V\textsubscript{rms} is more representative.\par}
\end{table}

Finally, Tab.~\ref{tab:tabla_lp_mode} and Tab.~\ref{tab:tabla_hp_mode}  summarises the simulation results of the presented implementation. The new LNA shows increased current for low-power (LP) mode and modest low-frequency  noise. The current penalty due to single-fault detection mechanism is approximately 2\,\textmu A at typical corner, which is included in the results of Tab.~\ref{tab:tabla_lp_mode} and Tab.~\ref{tab:tabla_hp_mode}. The high-performance, offset-cancellation (HP) mode consumes higher current at the cost of slightly increased noise in order to reject DC offset and low-frequency components. The HF noise $\sigma$ is high due to some extreme outliers. For a better indication, its median is around 11\,\textmu V\textsubscript{rms}.

\section{Conclusion} \label{sec:conclusion}

Recording-side patient safety in DC-coupled front ends has gone almost unaddressed in the circuits literature, even as stimulation safety has become well established. This work addresses that gap directly, with a single-fault detection mechanism that senses the DC imbalance of an LNA input-device failure and disables the amplifier before the resulting fault current can irreversibly damage cortical tissue. Implemented in 65\,nm CMOS, it resolves a worst-case fault of 6.4\,nA across all corners, the current at which an 8533\,\textmu m\textsuperscript{2} SIROF electrode reaches the ISO 14708-3 Clause 16.2 limit of 0.75\,\textmu A/mm\textsuperscript{2}.

The mechanism guards specifically against differential-pair failure. It does not cover electrostatic-discharge faults at the IO pad or the secondary ESD device, either of which can open its own DC path to the brain that a recording-side current comparison cannot observe. Future work should therefore target human-safe ESD protection whose failure modes do not endanger the patient.

\vfill\null

\bibliographystyle{IEEEtran}
\bibliography{IEEEabrv,Section/references}

\end{document}